# Enhanced Infrared Photodiodes Based on PbS/PbCl$_x$ Core/Shell Nanocrystals

*Adam E. Colbert,* Diogenes Placencia, Erin L. Ratcliff,* Janice E. Boercker, Paul Lee, Edward H. Aifer, and Joseph G. Tischler*

Dr. A. E. Colbert, Dr. D. Placencia, Dr. J. E. Boercker, Dr. E. H. Aifer, Dr. J. G. Tischler
U.S. Naval Research Laboratory
4555 Overlook Ave SW, Washington, D.C. 20375, USA
E-mail: adam.colbert.ctr@nrl.navy.mil

Prof. E. L. Ratcliff
Department of Chemical and Environmental Engineering, University of Arizona, 1133 E. James. E. Rogers Way, Tucson, Arizona 85721, USA
Department of Materials Science & Engineering, University of Arizona, 1235 East James E. Rogers Way, Tucson, Arizona 85721, USA
Email: ratcliff@arizona.edu

Prof. E. L. Ratcliff, P. Lee
Chemistry and Biochemistry Department, University of Arizona, 1306 East University Blvd, Tucson, AZ 85721, USA

Improved passivation strategies to address the more complex surface structure of large-diameter nanocrystals are critical to the advancement of infrared photodetectors based on colloidal PbS. In this contribution, the performance of short-wave infrared (SWIR) photodiodes fabricated with PbS/PbCl$_x$ (core/shell) nanocrystals versus their PbS-only (core) counterparts are directly compared. Despite their inherently similar bulk properties, devices using PbS cores suffer from shunting and inefficient charge extraction, while core/shell-based devices exhibit greater external quantum efficiencies and lower dark current densities. To elucidate the implications of the shell chemistry on device performance, the thickness-dependent energy level offsets and interfacial chemistry of the nanocrystal films with the zinc oxide electron-transport layer are evaluated. The disparate device performance between the two synthetic methods is attributed to unfavorable interface dipole formation and surface defect states, associated with inadequate removal of the native organic ligands in the core-only films. The core/shell system offers a promising route to manage the additional nonpolar

(100) surface facets of larger nanocrystals that conventional halide ligand treatments fail to sufficiently passivate.

## 1. Introduction

Photodetectors operating in the short-wave infrared (SWIR) region have numerous applications in degraded visual environments (DVE), biomedical imaging, industrial defect analysis, light detection and ranging (LiDAR), and telecommunications.[1, 2] Mature infrared detector technologies based on epitaxially grown semiconductors (InGaAs, Ge, InSb) are hindered by factors such as high-temperature processing, lattice matching constraints, incompatibility with flexible substrates, and the requirement for a hybridization process to be integrated with silicon Readout Integrated Circuits (ROICs). These constraints lead to high manufacturing costs, impose limits on pixel size (> 5 μm) and in turn, impact upon sensor resolution. Alternatively, colloidal semiconductor nanocrystals offer a promising route to realize wide-scale integration of infrared photodetectors due to their compatibility with low-cost solution processing techniques, flexible substrates, and direct deposition onto ROIC surfaces, allowing for the fabrication of monolithic high-resolution imagers.[3, 4]

Of the available materials options for colloidal systems, PbS nanocrystals have attracted considerable attention for infrared and multispectral detectors. Materials advantages include facile hot-injection synthesis, relatively high carrier mobilities, and a large Bohr exciton radius (~18 nm) that enables band gap tuning through particle size (0.6-1.6 eV or ~ 0.8-2.0 μm).[1, 5] Recent efforts have extended the sensitivity of PbS nanocrystal detectors into the mid-wave infrared (MWIR) region via a modified synthetic protocol which synthesizes large (>10 nm) particle sizes that approach the bulk band gap detection limit of 0.41 eV (~3 μm).[6] Long-wave infrared detectors have also been realized using sub-gap detection via intraband transitions in highly n-doped films.[5] PbS nanocrystals have been successfully integrated onto silicon ROICs,[4, 7-9] and subsequently implemented in commercial SWIR cameras.[10]

Infrared photodetectors based on PbS nanocrystals have achieved notable performance metrics in both photoconductor[6, 11] and photodiode[12-14] devices, reaching reported detectivity values that rival commercial InGaAs detectors.[15] However, many of these studies build on ligand exchanges and device architectures developed for photovoltaic devices with band gaps of approximately 1.3 eV (~950 nm).[16] As PbS nanocrystals increase in size, they exhibit a greater proportion of nonpolar (100) facets relative to cation-rich (111) facets.[17, 18] A weaker binding affinity of the native oleic acid surface ligands to the (100) facets and the formation of deep trap states in the presence of oxygen lead to a significant reduction of air stability with particle diameters exceeding roughly 4 nm.[17, 19] These factors have motivated the use of mixed ligand exchanges to provide more extensive surface passivation of narrow band gap nanocrystals necessary for SWIR devices.[12, 20, 21] Additionally, the nanocrystal band edge energies are altered both by particle size and ligand-induced surface dipole effects,[22] requiring more specific tuning of the device's electron- and hole-selective contacts to reduce barriers for charge extraction, while minimizing device dark currents and the subsequent detector noise.

To date, PbS optoelectronic device studies have primarily employed synthetic methods in which PbO is used as the lead precursor via the formation of lead-oleate.[23] The most common alternative synthesis uses a $PbCl_2$ suspension in oleylamine as the lead precursor.[24-26] We,[27] and others,[28] have recently found that this method produces particles with a $PbCl_x$ shell, resulting in higher photoluminescence quantum yields and superior resistance to oxidation than their counterparts synthesized with PbO precursors.[26-28] Despite the apparent advantages, $PbCl_2$-synthesized nanocrystals have seen little implementation in device studies[29] and the $PbCl_x$ shell has been suggested to form an insulating barrier for charge transport that would hinder their use in optoelectronic device applications.[28] Notably, Xia et al. recently demonstrated high-efficiency photovoltaic devices using 0.95 eV (~1.3 μm) band gap PbS with native chloride passivation, derived through cation-exchange of ZnS nanorods.[30] In this work, we investigate the influence of the $PbCl_x$ shell on the electronic properties of SWIR photodiodes. We compare devices with photosensitivity approaching 1.5 μm, fabricated with PbS nanocrystals derived from PbO (herein denoted as "core") and $PbCl_2$ (herein denoted as "core/shell") methods. Contrary to previous suggestions, devices using core/shell nanocrystals exhibit lower dark current densities, improved diode rectification, and higher photoresponse in the SWIR regime, when compared to core-only devices. Additionally, we use thickness-dependent X-ray/ultraviolet photoelectron spectroscopy (XPS/UPS) measurements to determine band-edge offsets at the interface of the core and core/shell nanocrystals with the ZnO electron transport layer. We find significant differences in the relative changes to the local vacuum levels between these two synthetic methods as a function of film thickness. This effect is attributed to interface dipoles that arise as the thin film is formed. Our results suggest that besides reduction of non-radiative mechanisms, the observed disparities in device performance are associated with dipole formation at the interface of the PbS nanocrystals with the ZnO layer resulting in a charge-extraction barrier, further clarifying prior reports of decreased charge transport in the core/shell system. We relate these differences to incomplete removal of the native surface ligands in the core-only films using established halide ligand exchange methods and demonstrate the utility of the core/shell synthesis to passivate the sensitive (100) surface facets of large-diameter particles.

## 2. Results and Discussion

### 2.1. Device Fabrication and Current-Voltage Characterization

PbS devices were fabricated based on the planar heterojunction device architecture reported by Chuang et al.[16] that has served as the basis for record efficiency PbS nanocrystal photovoltaic devices[31, 32] (**Figure 1**a, inset). The electron-selective contact is a ZnO nanocrystal film on a patterned ITO-coated glass substrate, upon which we deposit the PbS absorber layer via spin-coating in a layer-by-layer process. The bulk of the film (≥ 240 nm) is ligand exchanged with tetrabutylammonium iodide (TBAI), topped with a thin (40 nm) 1,2-ethanedithiol (EDT)-treated layer to facilitate hole extraction and reduce recombination at the hole-collecting electrode interface.[16, 22] The back contacts are subsequently deposited via thermal evaporation of $MoO_x$ (15 nm) followed by gold (100 nm). Insertion of a thin $MoO_x$ interlayer between the PbS film and the metal electrode is used to improve current extraction and diode rectification by preventing formation of a Schottky barrier at the contact interface.[33] Thus, the $MoO_x$ layer facilitates hole extraction and interfacial energetic alignment through midgap states in the oxide[34] that are formed via reduction of the $MoO_x$ in chemical redox reactions with the underlying PbS-EDT layer.[35]

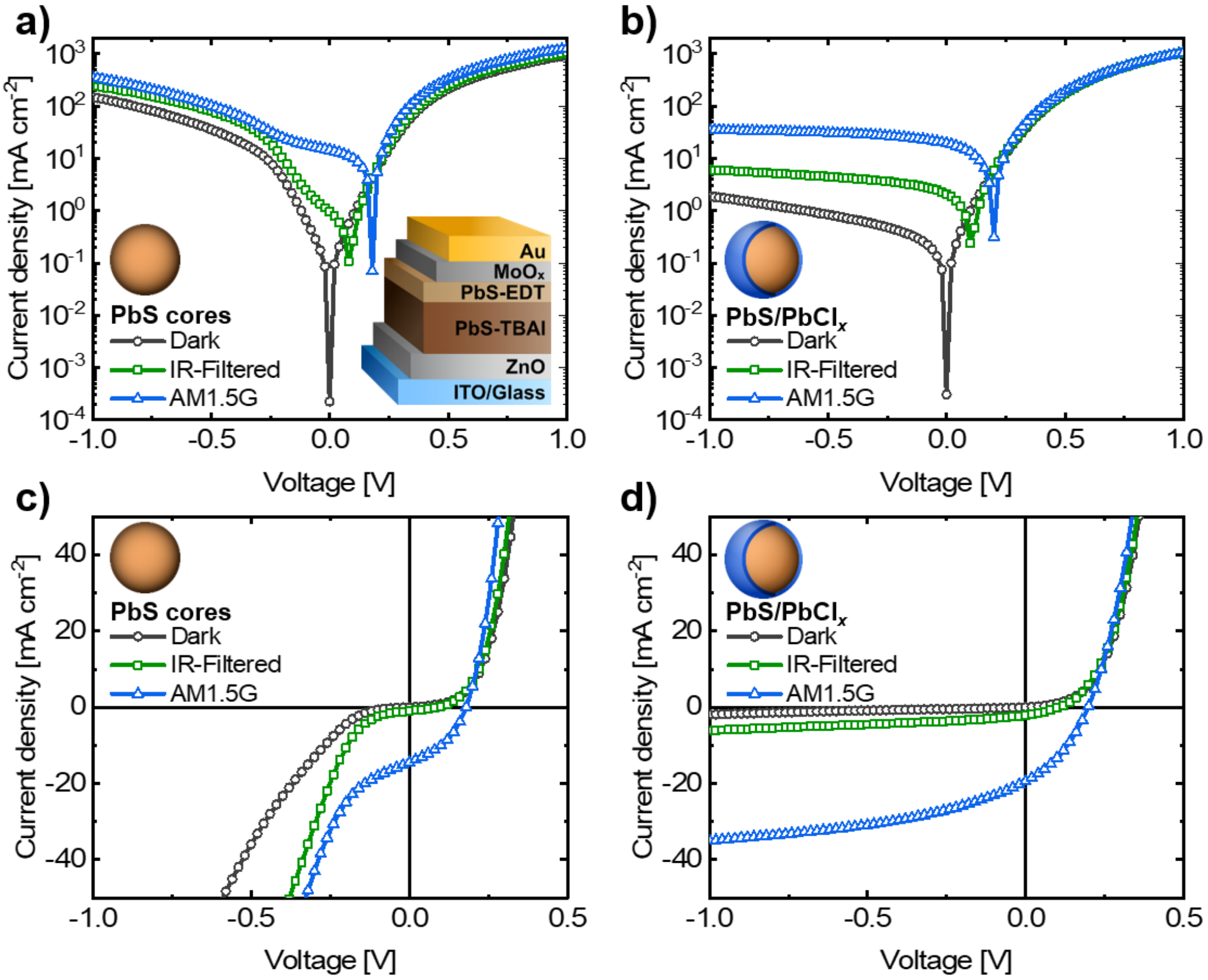

**Figure 1.** Semilog (top panel) and linear (bottom panel) scale *J*-*V* measurements for photodiodes with a,c) core and b,d) core/shell absorber layers, respectively, where the device active areas are 0.019 $cm^2$. The inset in a) depicts the device structure.

**Table 1.** Photovoltaic performance parameters for core and core/shell devices.[a)]

| PbS Type | Illumination | $J_{SC}$ [mA cm$^{-2}$] | $V_{OC}$ [V] | FF [%] | $R_{SH}$ [Ω cm$^2$] | $R_S$ [Ω cm$^2$] |
|---|---|---|---|---|---|---|
| Core | AM1.5G | -14.5 ± 0.9 | 0.18 ± 0.002 | 39 ± 0.6 | 29 ± 0.5 | 2.5 ± 0.3 |
| Core/Shell | AM1.5G | -20.1 ± 0.5 | 0.18 ± 0.002 | 36 ± 0.8 | 27 ± 1.6 | 3.3 ± 0.2 |
| Core | IR | -1.0 ± 0.06 | 0.10 ± 0.002 | 29 ± 0.3 | 93 ± 3.1 | 28 ± 2.5 |
| Core/Shell | IR | -2.1 ± 0.05 | 0.11 ± 0.001 | 33 ± 0.3 | 114 ± 4.6 | 17 ± 0.6 |

[a)]The reported values are the averages from 21 devices with 0.019 cm$^2$ active areas, and error bar uncertainties indicate their 95% confidence interval.

Figure 1 compares the current density-voltage (*J-V*) characteristics of core (Figure 1a,c) and core/shell (Figure 1b,d) infrared photodiodes under dark conditions and Air Mass 1.5 Global (AM1.5G) simulated solar illumination. Infrared illumination was provided via filtering of the AM1.5G output through a 1100 nm (c-silicon) long-pass filter, which is also included. For reference, both the core and core/shell nanocrystal absorption spectra are shown against the AM1.5G solar spectrum in **Figure S1** (Supporting Information), in which the c-Si band edge is demarcated. The device performance parameters are summarized in **Table 1**, and the error bars for the *J-V* curves are shown in **Figure S2** (Supporting Information). Under broadband AM1.5G illumination, the short-circuit current density ($J_{SC}$) is 28% greater in the core/shell devices than the core devices. In both cases, we observe photoresponse under infrared illumination, with the core/shell devices generating approximately double the $J_{SC}$ of the core devices. However, the open-circuit voltages ($V_{OC}$) are consistent at 0.18 V (0.1 V), under AM1.5G (IR-filtered) illumination, and fill factors (*FF*) differ by only 3-4% in both cases. We note that the series ($R_S$) and shunt ($R_{SH}$) resistance both increase under IR exposure, compared to AM1.5G illumination. We attribute this effect to the lack of UV activation of ZnO in the former, which is thought to promote the desorption of surface bound oxygen molecules on the ZnO, thereby increasing the n-doping and conductivity of the film.[36, 37] While beyond the scope of this work, various surface treatments[38] and doping strategies[39] have been developed to address this issue in PbS optoelectronic devices. Taken together, this data indicates that the core/shell synthetic method shows a notable improvement in device performance relative to core route, especially with regards to photodetector-relevant factors; namely, low dark current and high $J_{SC}$.

In Figure 1a,c we find the core devices show low-performing diode rectification in all cases, with significant shunting observed when the applied reverse bias exceeds -200 mV, where increased leakage current is readily apparent. This deviates from the flat reverse bias behavior observed in previous works with PbS core devices using the same architecture with wider band gap (1.2 eV) nanocrystals,[35] as well as alternative device structures using narrow band gap PbS with organic thiol ligand treatments.[13] In contrast, the core/shell devices show significantly reduced dark current densities in reverse bias and improved diode rectification characteristics (Figure 1b,d) with rectification ratios of ≈1 x $10^3$ (±1 V), approximately two orders of magnitude greater than the core devices. We compare the dark current densities ($J_{dark}$) and differential resistance ($\Delta V/\Delta J$) of the dark *J-V* curves for these devices at zero-bias and at low reverse bias in **Table S1** (Supporting Information), where a better photodiode will have lower $J_{dark}$ and higher resistance.[40] At 0 V, we find both the $J_{dark}$ (≈ -3 x $10^{-4}$ mA $cm^{-2}$) and the shunt resistance (≈ 300 Ω $cm^2$) are similar between the two materials. However, under reverse bias the $J_{dark}$ of the core devices are 10-fold greater than the core/shell devices at -0.25 V (-8.4 vs -0.48 mA $cm^{-2}$), and 100-fold greater at -0.5 V (-36 vs -0.84 mA $cm^{-2}$). Correspondingly, the differential resistance of the core devices over the same voltage range falls under 10 Ω $cm^2$, while the core/shell devices remain greater than 600 Ω $cm^2$, indicating a significantly larger resistance to leakage current, possibly in the form of an energetic barrier.

### 2.2 External Quantum Efficiency Characterization

The device external quantum efficiency (EQE) spectra at zero-bias in the near-infrared (NIR)/SWIR region are shown in **Figure 2**, and the performance in the infrared region is summarized in **Table 2**. We perform these measurements with larger active area (0.125 $cm^2$) devices measured through a mask to minimize additional current collection associated with edge effects in smaller area devices.[41] Additionally, we investigate the influence of increasing the PbS layer thicknesses from 260 nm (corresponding to the *J-V* data in Figure 1) to 340 nm, to increase light absorption. The corresponding responsivity spectra (**Figure S3,** Supporting Information) and the full-range EQE spectra with the associated dark *J-V* curves are given in **Figure S4** and **Figure S5** (Supporting Information). In accordance with the data discussed above, the $J_{dark}$ of both devices at 0 V is on the order of 1 x $10^{-4}$ mA $cm^{-2}$, and increases at -0.5 V to approximately -10 and -0.3 mA $cm^{-2}$ for the core and core/shell devices, respectively. Similarly, the differential resistances of the core devices at -0.5 V are 100-fold lower than the core/shell devices, demonstrating the shunting behavior under reverse bias.

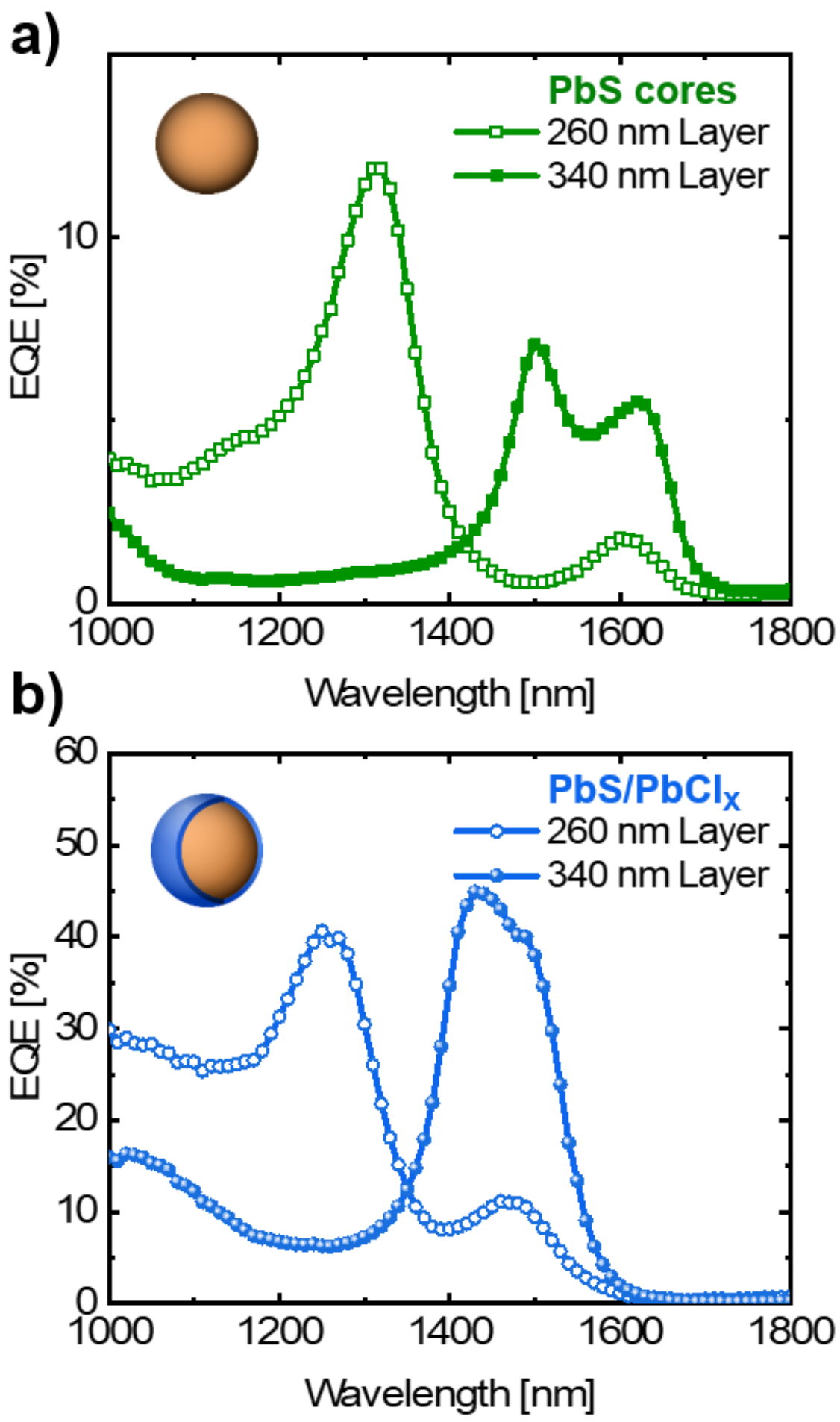

**Figure 2.** EQE spectra at 0 V bias for a) core-only and b) core/shell nanocrystal devices, with active layer thickness of 260 nm (open symbols) and 340 nm (closed symbols). All devices were fabricated with a 0.125 $cm^2$ active area.

**Table 2.** Infrared performance and dark current characteristics for core and core/shell devices.

| | | Maximum IR Performance[a)] | | | 0 V Applied Bias | | -0.5 V Applied Bias | |
|---|---|---|---|---|---|---|---|---|
| PbS Type | Thickness [nm] | $\lambda_{max}$ [nm] | EQE [%] | Responsivity [A W$^{-1}$] | $J_{dark}$ [mA cm$^{-2}$] | $\Delta V/\Delta J$ [Ω cm$^2$] | $J_{dark}$ [mA cm$^{-2}$] | $\Delta V/\Delta J$ [Ω cm$^2$] |
| Core | 260 | 1320 | 12 | 0.13 | $-7.6 \times 10^{-5}$ | 700 | -9.4 | 20 |
| Core | 340 | 1500 | 7 | 0.09 | $-1.9 \times 10^{-4}$ | 210 | -7.4 | 30 |
| Core/Shell | 260 | 1250 | 41 | 0.41 | $-1.0 \times 10^{-4}$ | 600 | -0.28 | 1800 |
| Core/Shell | 340 | 1430 | 45 | 0.52 | $-8.7 \times 10^{-5}$ | 760 | -0.31 | 1900 |

[a)]The maximum detector performance parameters for $\lambda > 1000$ nm for each PbS device from Figure 2 (0.125 cm$^2$ active areas).

Comparing the EQE spectra from devices of different active layer thickness, we observe the presence of an intrinsic optical cavity effect,[42] with apparent enhancement in the 1.2-1.3 μm region for the 260 nm-thick devices (Figure 2a,b, open symbols), and in the 1.4-1.5 μm region in the case of the 340 nm-thick devices (Figure 2a,b, closed symbols) for both core and core/shell devices. This Fabry-Perot resonance effect has been observed previously in SWIR PbS photodiodes with similar device structures.[42, 43] The EQE of the core devices remain primarily under 10%, with maximum values of 12% at 1.3 μm for the 260 nm-thick devices, and 7% at 1.5 μm for the 340 nm-thick devices (Figure 2a). The core/shell system shows improved performance, with maximum EQE values of 41% at 1.25 μm for the 260 nm-thick devices, and 45% at 1.43 μm for 340 nm-thick devices (Figure 2b). This is consistent with the trend observed with the overall greater $J_{SC}$ values in the case of core/shell devices under filtered illumination in the NIR/SWIR region. However, we note that there were discrepancies in the integrated product of the EQE spectra with the AM1.5G spectrum to calculate the $J_{SC}$ of these devices for comparison to the measured values from Table 1, summarized in the **Table S2** (Supporting Information). In the case of the core devices, the EQE predicts approximately half the measured $J_{SC}$ value (-7 mA cm$^{-2}$ calculated vs -14.5 mA cm$^{-2}$ measured), while the core/shell values agree well (-22 mA cm$^{-2}$ calculated vs -20 mA cm$^{-2}$ measured). While we are unable to identify the source of this discrepancy in the core device measurements, it may result from higher levels of shunting due to film defects in the core-only films, such as pinholes or nonidealities in the PbS surfaces associated with midgap states,[44] that would lead to greater losses in carrier collection in the larger area devices used for the EQE measurements (0.125 $cm^2$) than with the smaller area devices used for the *J*-*V* characterization (0.019 $cm^2$).[41] Nevertheless, the overall trend of greater optical response in the SWIR region with core/shell devices remain consistent.

There are several possible explanations for the differences in device performance between the two photo-active layers. One consideration is that differences in interfacial recombination due to back transfer of electrons at the ZnO contact may factor into the $J_{SC}$ losses in the core devices.[45] However, this is not a dominant factor, as surface recombination would not lead to the observed shunting and diode breakdown characteristics under reverse bias. A second, and more likely explanation, is that as the electric field across the PbS film increases with greater reverse bias, breakdown can occur via carrier tunneling across the depletion region, leading to an avalanche-type of breakdown at higher reverse bias.[44] The observed breakdown characteristics of the core devices could therefore be caused by the lack of a sufficiently thick depletion width in the PbS film. In the case of the core/shell devices, the improved diode rectification suggests a thicker depletion width and/or the formation of an interface dipole that opposes parasitic current flow under reverse bias conditions. An additional third and also likely consideration is that a high prevalence of interfacial states at the PbS/ZnO heterojunction[36] can result in pathways that enable carrier transport in the reverse direction via defect-mediated tunneling,[46] leading to shunting and reduced breakdown voltage under reverse bias. These mid-gap states are typically attributed to surface defects induced by oxidation, inadequate surface passivation such as undercoordinated surface sites, or incomplete exchange of the native ligands.[19] Importantly, the native halide passivation and resistance to oxidation provided by the $PbCl_x$ shell will reduce the presence of the mid-gap states that facilitate defect-mediated tunneling effects in the core devices. We note however that in all three outlined possibilities, the differences in surface chemistry from the $PbCl_x$ shell is postulated to influence the nanocrystal energy levels through surface dipole effects.[22]

### 2.3. Thickness-Dependent Interfacial Energy Level and Chemical Analysis

Our device measurements suggest that the disparities between the core and core/shell absorber layers may stem from differences in interfacial trap states, depletion widths, and/or interface dipoles at the PbS/ZnO interface. To further understand the physical and chemical mechanisms resulting in the observed differences, we performed thickness-dependent XPS/UPS measurements of PbS nanocrystals upon ZnO surfaces. **Figure 3**a,b shows the UPS spectra for a series of sequentially-deposited core and core/shell nanocrystals upon a ZnO surface. The secondary edge kinetic energy in Figure 3a has been converted to work function ($\Phi$), the energy difference between surface vacuum level and Fermi level, and the valence band region is displayed as the energy with respect to the Fermi level in Figure 3b. Semiquantitative band-edge offsets (Figure 3c), as a function of nanocrystal film thickness, were obtained from data in Figure 3a,b. The optical gaps for both the PbS and ZnO films denoted in Figure 3c were obtained from UV/Vis/NIR measurements (Figure S1, Supporting Information).

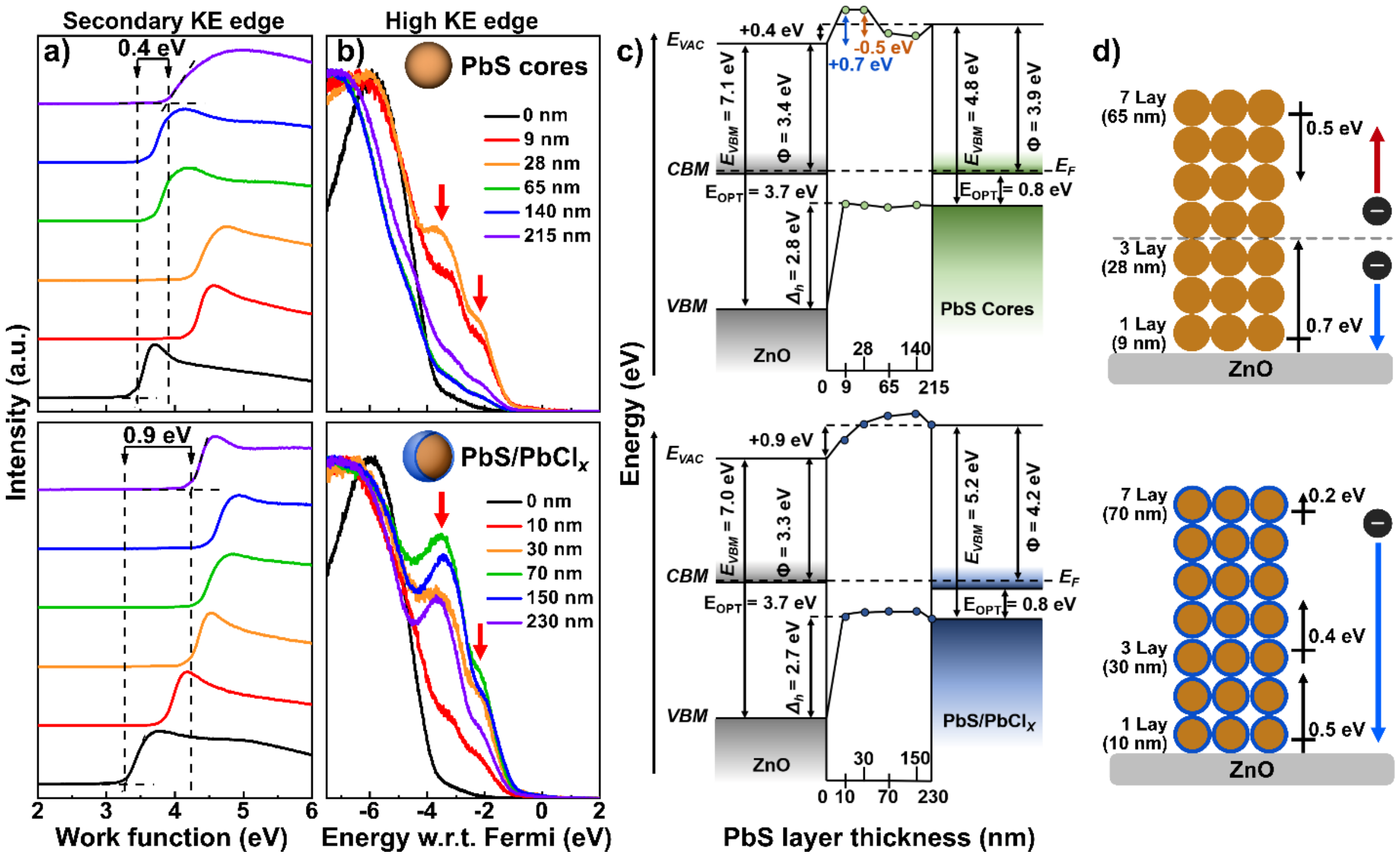

**Figure 3**. a,b) UPS spectra for the successive deposition of core-only (top panel) and core/shell (bottom panel) nanocrystals. The secondary kinetic energy edges in a) and high kinetic energy edges in b) are plotted with respect to the surface work function and system Fermi level, respectively. The surface work functions were obtained using the intersection of the dashed tangent lines, with the initial and final values indicated by the vertical dashed lines. The red arrows in b) indicate sulfur 3p levels. c) Nanocrystal band-edge offsets at the ZnO/PbS interface for the work function and $E_{VBM}$ positions as a function of PbS layer thickness in core (top panel) and core/shell (bottom panel) films. d) Diagram depicting the vacuum level shifts measured as a function of film thickness for the first ~70 nm in PbS core (top panel) and core/shell (bottom panel) films deposited on ZnO, with the direction of the associated surface dipole indicated. The number of layers here indicates the total number of individual PbS depositions for each of the first 3 XPS/UPS measurements of PbS atop ZnO.

In Figure 3a, the bare ZnO surfaces reflect low work functions (3.3-3.4 eV), common amongst the spun cast zinc-based metal oxide systems.[47] Tail states in the valence band region (~3.7 eV from Fermi; Figure 3b) of both films are also noticeable and characteristic.[47] Upon the deposition of the first layer (9-10 nm), both systems immediately display the characteristic sulfur 3p levels near the Fermi level, as described previously by Miller et al. and indicated by the red arrows in Figure 3b.[48] At this point, determination of an initial valence band maximum (VBM) is possible by finding the intersection between the spectrum baseline and the rise of the peak in closest proximity to the Fermi level, while applying a 0.2 eV correction factor due to the low density of states at the VBM for the nanocrystals.[48] Simultaneously, a dramatic shift of the secondary kinetic energy edge towards a higher work function is noted for both systems, indicative of a net positive charge present at the surface, thereby increasing the local surface vacuum level (see further below). Previous work has shown that these initial energy redistribution characteristics are indicative of localized redox-type processes that serve to maintain a charge balance within the system, which can result in the creation of interface dipoles and/or band bending phenomena.[35] The subsequent deposition of PbS atop of the already deposited film pronounces the sulfur orbitals near the Fermi level even further, while changes in the work function are also noticeable until the deposition of the final layer.

It is important to note that despite their initially similar characteristics, the core and core/shell systems have two large distinctions. First, upon deposition of the first layer of nanoparticles (~10 nm), we observe a difference in the near surface vacuum level shift. Both the core and the core/shell systems show an upwards shift in the surface vacuum level leading to an increase in work function to 4.2 eV and 3.8 eV, respectively. This +0.7 eV shift for the core and +0.5 eV shift for the core/shell most possibly derives from a mixture of interface dipoles and band bending. XPS of the Zn $2p_{3/2}$ core level before and after the initial deposited layer (**Figure S6**, Supporting Information) shows a slight shift (~0.2 eV) towards the Fermi level, indicative of a slight upward shift in the conduction band of ZnO. The larger net work function shift at the ZnO/core interface would exhibit a higher barrier to extraction and/or possible tunneling. However, due to the thickness of the initial film (~10 nm), in addition to the escape depth of core electrons being ~5-10 nanometers below the surface in XPS, dipole effects (~1-2 nm) cannot be separated directly from band bending, in addition to depletion layer formation directly at the ZnO/PbS interface.

A second notable observation is the change in the surface work function between the second and third measured layers (i.e., between ~30 nm and ~70 nm). In the core system, we observe an inversion in the surface dipole, with the work function decreasing from 4.2 to 3.7 eV. This is further supported by the UPS valence band spectrum (**Figure S7**a, Supporting Information) with the apparent change in the density of states profile, indicated by the loss of the S 3p peaks between the second (~30 nm-thick) and third (~70 nm-thick) measurements. No differences in the Pb 4f and S 2p core levels were observed (Figure S7b,c, Supporting Information), indicating the surface of the particles were not oxidized. Alternatively, in the core/shell system, we detected a continual increase in the near surface work function, with a value of 4.2 eV for 30 nm, 4.4 eV for 70 nm, and 4.5 eV for 150 nm thickness, resulting in an overall shift of approximately 1.2 eV. No discernible changes were detected in the valence band region or the Pb 4f and S 2p cores (**Figure S8**, Supporting Information). The differences in surface dipole shifts between the core and core/shell systems near the ZnO interface are depicted as a function of film thickness in Figure 3d.

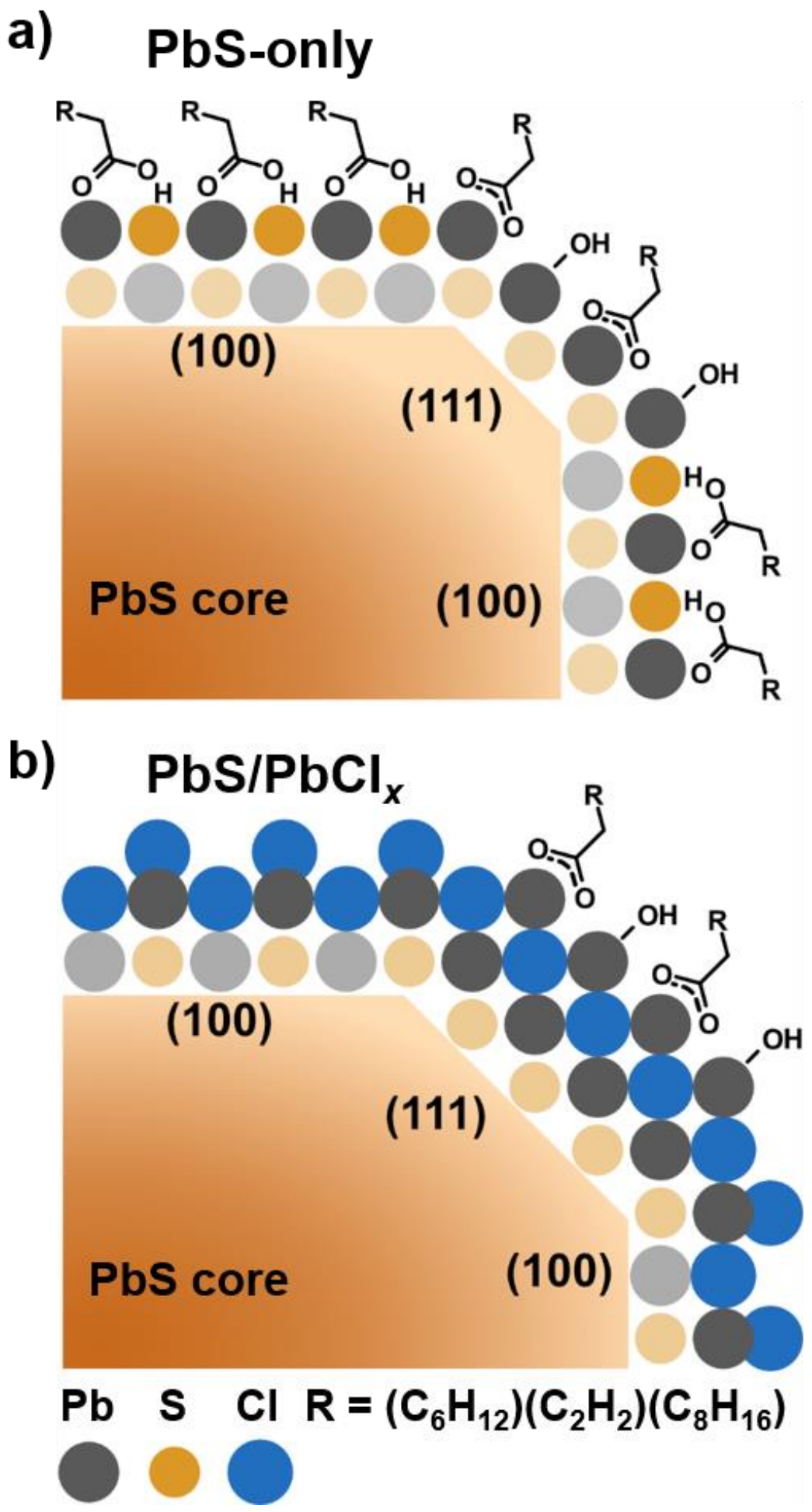

**Figure 4.** Simplified cross-sectional illustration representing the native surface chemistry on the (100) and (111) facets of a) core and b) core/shell PbS, based on the surface structures proposed in Ref. 17 and 28, respectively. The full three-dimensional structures, together with the surface ligands, are charge-neutral in both cases.

We rationalize the dramatic difference in the surface vacuum levels as being due to charge trap states associated with inadequate ligand exchange in the cores, as observed by differences in the O 1s and C 1s spectra (**Figure S9**, Supporting Information), relative to the core/shell films (**Figure S10**, Supporting Information) that show near complete removal of the native ligands. We emphasize that incomplete ligand exchange and the associated surface dipole inversion discussed above would result in differences in the depletion widths and/or tunneling barrier effects described in the device analysis. Our findings can be explained by the increased presence of stoichiometric (100) crystal facets relative to cation-rich (111) facets as the nanocrystal size increases.[18] Halide surface treatments readily exchange the anionic oleate and hydroxyl ligands that coordinate the polar (111) facets, but exhibit reduced affinity for the nonpolar (100) facets (**Figure 4**a).[17, 21] Instead, the low surface energy of the (100) facets enables removal of the weakly-bound oleic acid ligands in the presence of polar solvents such as methanol, allowing the exposed facets to epitaxially fuse.[49-51] Balazs et al. determined that solid-state halide ligand exchange is acid-catalyzed, and the reactivity of the ligands can be tuned by the acidity of the solution to influence the degree of ordering in the film and the resulting electronic properties.[49] The authors further demonstrate that TBAI/methanol treatment of 3.8 nm diameter PbS yields incomplete removal of oleic acid ligands, leading to high trap state densities and low conductivity in field effect transistor measurements.[49] The surface structure of the PbS/$PbCl_x$ nanocrystals proposed by Winslow et al. is depicted in Figure 4b, in which the (111) facets of the core/shell PbS have a monolayer of chlorine atoms substituting the lattice sites that would be occupied by sulfur if the nanocrystal grew larger, underlying a terminal lead cation layer that is passivated by the solubilizing oleate ligands.[28] In this description, the (100) facets have $PbCl_2$ termination without additional coordinating ligands, therefore circumventing the need to separately address passivation of those surfaces.[28] The differences in native surface chemistry between the core and core/shell systems are depicted in Figure 4.

Finally, we relate our findings to recent PbS core device studies that employ solid-state ligand exchange with a mixture of zinc iodide ($ZnI_2$) and 3-mercaptopropionic acid (MPA) in methanol, using similar device structure and fabrication methods to those reported here. Devices with 1.35 μm band gap PbS show a significant performance enhancement with the $ZnI_2$/MPA treatment compared to iodide-only treatment, achieving $J_{SC}$ values of 26.6 and 11.8 mA cm$^{-2}$, respectively under AM1.5G illumination.[12] This improvement is considerably more pronounced than demonstrated in their previous work with smaller (0.95 μm band gap) nanocrystals with $J_{SC}$ of 24.4 and 19.5 mA cm$^{-2}$ for $ZnI_2$/MPA vs iodide-only treatments, respectively.[52] We posit that the lower steric hindrance of the $Zn^{2+}$ cation[49] and the additional thiol and carboxylic acid functional groups offered by MPA enable a more complete ligand exchange and improved passivation of the (100) facets, contributing to this discrepancy in performance enhancement with $ZnI_2$/MPA treatment between large and small PbS devices. Furthermore, the $J_{SC}$ reduction between small[52] and large[12] PbS cores using iodide-only treatment is consistent with our observations using smaller (1 μm band gap) cores[35] and the data presented here.

## 3. Conclusion

In summary, we have demonstrated the use of PbS/$PbCl_x$ core/shell nanocrystals in SWIR photodiodes, with sensitivity ranging from the UV region to 1.5 μm. Core/shell-based devices exhibit greater external quantum efficiencies and significantly reduced dark current densities in reverse bias, as compared to conventional PbS cores, allowing greater overall signal-to-noise detection. Investigation of the PbS/ZnO interface using thickness-dependent XPS/UPS measurements indicate a greater shift of the local vacuum level in the case of the core/shell system, resulting in an effective 1.2 eV shift across 150 nm, attributed to differences in the interface dipoles. Our measurements suggest that mid-gap states associated with inadequate ligand exchange in the core devices may enable defect-mediated tunneling and lead to unfavorable dipole formation to oppose parasitic current flow under reverse bias conditions. In contrast, the core/shell PbS show near complete removal of the organic surface ligands and stronger interface dipole formation that promotes electron collection and mitigates losses due to carrier recombination, resulting in the observed improvements in device performance. Our findings affirm the viability of core/shell nanocrystals for optoelectronic device applications and will help inform the optimization of PbS-based SWIR devices in future studies.

## 4. Experimental Section

*General Methods*: All manipulations were performed using standard air-free techniques on a Schlenk line or in a nitrogen-filled glove box, unless otherwise stated.

*Materials*: All chemicals were used without further purification. Lead(II) oxide (PbO, 99.999%), oleic acid (OA, 90%), 1-octadecene (ODE, 90%), bis(trimethylsilyl)sulfide (TMS, synthesis grade), lead(II) chloride ($PbCl_2$, 98%), sulfur (≥ 99.99%), oleylamine (tech. grade, 70%), zinc acetate dihydrate (99.999%), potassium hydroxide (KOH, ≥85%), tetrabutylammonium iodide (TBAI, 98%), 1,2-ethanedithiol (EDT, ≥98.0%), hexane (anhydrous, 95%), heptane (anhydrous, 99%), toluene (99.8%), acetonitrile (anhydrous, 99.8%), methanol (anhydrous, 99.8%), ethanol (200 proof, anhydrous, ≥99.5%), Triton X-100 (lab grade), gold coated microscope slides (layer thickness 1000 Å, 99.999% Au) were obtained from Sigma-Aldrich. Molybdenum(VI) oxide (99.9995%) was obtained from Alfa Aesar. Gold shot (99.99%) was obtained from R.D. Mathis. S1813 photoresist was obtained from Rohm and Haas. Indium tin oxide (ITO)-coated glass (150 nm thick layer, sheet resistance of ~10 $\Omega$ $sq^{-1}$) was obtained from Delta Technologies Ltd.

*PbS Core Synthesis*: PbS nanocrystal cores with a first exciton transition energy centered at ~0.8 eV were synthesized following a modified version of the procedure developed by Lee et al., using standard air-free Schlenk techniques.[13] Stock solutions of

bis(trimethylsilyl)sulfide (TMS) precursors were prepared in dry ODE at 95 mM (120 μL TMS in 6 mL ODE) and 39.5 mM (150 μL in 18 mL ODE). PbO (0.892 g, 4 mmol), ODE (100 mL) and oleic acid (7.6 mL) were combined in a 250 mL 3-neck flask and degassed at 110° C for 60 min. The flask was transitioned to argon atmosphere and the temperature reduced to 100° C. Nanocrystal nucleation was initiated by rapidly injecting 6 mL of 95 mM TMS precursor solution. Three subsequent injections were performed at 10 min. intervals from the initial injection using 6 mL aliquots of 39.5 mM stock TMS solution to grow the nanocrystals. 10 min. after the final injection, the flask was by rapidly cooled to 75° C, then slowly cooled to room temperature. Cleaning was performed in a nitrogen glove box with anhydrous solvents. The PbS product was isolated from the unreacted precursors and byproducts by diving evenly between eight centrifuge tubes, to which 19 mL toluene and a subsequent 12 mL acetonitrile were added. The particles were collected via centrifugation and the clear supernatant was discarded. The nanocrystal precipitate was dispersed in toluene and consolidated into two tubes, each with 10 mL, followed by precipitation with 12 mL acetonitrile and centrifugation. This cleaning step was performed a total of three times. Finally, the nanocrystals were dispersed in hexane, transferred to a clean vial, and dried under a stream of nitrogen in the glove box. The product was stored dry in the glove box until used.

*PbS Core/Shell Synthesis*: Core/shell nanocrystals with a first exciton transition energy centered at ~0.8 eV were synthesized following a modified version of the procedures from Weidman et al.[26] with a 12:1 Pb:S ratio, and scaled up by a factor of 10. The sulfur precursor was prepared in a Schlenk flask, to which sulfur powder (132 mg, 4 mmol) and dry oleylamine (5.5 mL) were added and briefly heated to 120° C under argon atmosphere with vigorous stirring, until the sulfur was completely dissolved. The flask was then removed from the heating mantle and slowly cooled to room temperature, while maintaining the inert atmosphere. $PbCl_2$ (12.5 g, 45 mmol) and oleylamine (75 mL) were mixed in a 250 mL 3-neck flask with continuous stirring, and degassed at 110° C for 30 min. The flask was then transitioned to argon atmosphere and the temperature increased to 120° C. Sulfur precursor (5 mL) was rapidly injected into the flask, and the mixture was left to stir at 120° C for 60 min., after which the reaction was quenched via rapid cooling to approximately 75° C, then slow cooling to room temperature. Cleaning steps were performed in a nitrogen glove box with anhydrous solvents, and all centrifugation steps were performed at 6000 rpm. First, 30 mL hexane was added to the flask and mixed. The product was split between four 50 mL centrifuge tubes, and excess $PbCl_2$ was removed from the suspension via centrifugation. The nanocrystal phase (suspended in hexane) was decanted into clean tubes, precipitated with 12 mL ethanol per tube, and centrifuged. The reddish-orange supernatant was discarded, and the nanocrystals in each tube were dispersed in 7 mL hexane. The tubes were again centrifuged to remove additional $PbCl_2$ (white precipitate) and the supernatant was decanted into two clean tubes. To each tube containing ~14 mL of PbS in hexane, 1 mL dry oleic acid was added, mixed well, and allowed to stand for 5 min. The particles were precipitated with 12 mL ethanol per tube and collected via centrifugation. One additional ligand exchange step was performed by suspending the precipitate in each tube with 10 mL hexane and 1 mL oleic acid. Excess oleic acid was removed by twice dispersing the precipitate in each tube with 10 mL toluene, precipitating with 12 mL acetonitrile, and centrifuging. The clean nanocrystal product was suspended in hexane, transferred to a vial, and dried under a stream of nitrogen in the glove box. The product was stored dry in the glove box until used.

*ZnO Nanocrystal Synthesis*: Colloidal ZnO nanocrystals were prepared in ambient conditions using the protocol from Chuang et al.[16] Zinc acetate dihydrate (2.95 g, 13.4 mmol) was dissolved in 125 mL methanol in an Erlenmeyer flask stirred at 60° C. A potassium hydroxide solution (0.4 M) was prepared by dissolving KOH (1.48 g, 26.4 mmol) in 65 mL methanol, and subsequently added dropwise to the zinc acetate solution via syringe pump at a rate of 3.25 mL $min^{-1}$. The reaction was left stirring at 60° C for 2.5 hours following the addition. The ZnO nanocrystals were collected via centrifugation and washed twice by dispersing in methanol. The final product was dispersed in 20 mL each of chloroform and butanol.

*PbS Nanocrystal Device Fabrication*: ITO substrates (1 in. x 1 in.) were cleaned by scrubbing with 5% Triton X-100, followed by sonication in 5% Triton X-100 (15 min.), nanopure water (5 min.), then absolute ethanol (15 min.). The ITO substrates were then patterned using a positive resist (S1813), processed with the necessary exposure, development, and heating steps. The exposed ITO was etched by submersion in TE-100 tin oxide etchant (5 min.) on a hotplate set to 120° C. The resist was removed via sonication in acetone (15 min.), and the substrates were again cleaned using steps listed above. PbS nanocrystal films were spin-cast in ambient conditions using a layer-by-layer approach based on previous literature reports.[16, 35] ZnO nanocrystals were deposited onto the ITO substrates by spin-coating two layers at 2000 rpm, followed by thermal annealing at 200° C for 10 min. PbS was dispersed in anhydrous n-heptane (20 mg $mL^{-1}$) and filtered to remove particulates (PTFE, 0.45 μm pore size). Each PbS layer was deposited by spin-coating at 2000 rpm, followed by the necessary ligand exchange steps, until the desired thickness was reached. The bulk of the film was ligand exchanged with tetrabutylammonium iodide (TBAI, 10 mg $mL^{-1}$ in methanol), rinsed twice with methanol and spun dry. The final two layers (~40 nm) were treated with 1,2-ethanedithiol (EDT, 0.05% v/v in acetonitrile) and rinsed twice with acetonitrile. The PbS films were air annealed at 90° C (3 min) and transferred to a nitrogen glove box with a built-in thermal evaporator, in which 15 nm $MoO_X$ and 100 nm gold films were successively deposited at a rate of 0.5 and 1 Å $s^{-1}$, respectively.

*Device Characterization*: Devices used for the current density-voltage (*J-V*) measurements were patterned with device areas of 0.019 $cm^2$, while those used for quantum efficiency measurements had active areas of 0.125 $cm^2$. For testing, the devices were contained in a hermetically sealed chamber with feedthroughs for the electrical contacts. Data for the *J-V* measurements was acquired using a Keithley 2400 source meter. Illumination was provided by a 1000 W solar simulator (Newport) using an Air Mass 1.5 Global (AM1.5G) filter, with the intensity calibrated to 100 mW $cm^{-2}$ using a standardized silicon reference cell. For the IR-illumination, the AM1.5G illumination was filtered using a c-silicon 1100 nm long-pass filter. Data for the external quantum efficiency measurements was collected on a home-built setup using illumination from a 150 W Xenon DC arc lamp (Newport) and a Cornerstone 130 monochromator (Newport), modulated with a SR 540 chopper (Stanford Research Systems) system at 250 Hz. Data was acquired with a SR560 low-noise preamplifier and a SR810 lock-in amplifier (Stanford Research Systems). The incident light intensity was determined using silicon (818-SL/DB) and germanium (818-IR/DB) NIST-traceable calibrated photodiodes (Newport).

*X-ray/Ultraviolet Photoelectron Spectroscopy*: Substrates were prepared by cleaving gold-coated glass microscope slides (100 nm layer thickness) with a titanium adhesion layer, and cleaning via sonication in 5% Triton X-100, nanopure water, then absolute ethanol. The ZnO layer was deposited by spin-coating two successive layers at 2000 rpm for 1 min. and annealing at 200° C for 15 min. Nanocrystal samples were dispersed in dry hexane at a (5 mg $mL^{-1}$). Each layer was deposited by spin-coating at 2000 rpm (30 s), treated with TBAI (10 mg $mL^{-1}$ in methanol) for 30 s, and spun dry. Residual ligands were removed by rinsing the surface with methanol and spinning dry three times. Characterization was performed with a Kratos Axis-Ultra using an Al Kα source (1486.6 eV) for XPS and a He(I) excitation source (21.2 eV) for UPS at a base pressure of $10^{-9}$ Torr. The samples were biased at -10.0 V to enhance collection of the lowest kinetic energy electrons during UPS experiments. All UPS spectra were referenced to the Fermi level ($E_F$) of a neat polycrystalline gold substrate.

**Supporting Information**
Supporting Information is available below.

**Acknowledgements**
The Office of Naval Research (ONR) is gratefully acknowledged for their financial support of this work. E.L.R. acknowledges the 2020 ONR Summer Faculty Research Program. A.E.C. acknowledges the National Research Council (NRC) postdoctoral program.

**Conflict of Interest**
The Naval Research Laboratory (NRL) has filed a provisional patent application related to this work. The inventors include: A.E.C., D.P., J.E.B., E.L.R., E.H.A., and J.G.T.

# Supporting Information

## Enhanced Infrared Photodiodes Based on PbS/PbCl$_x$ Core/Shell Nanocrystals

*Adam E. Colbert,* Diogenes Placencia, Erin L. Ratcliff,* Janice E. Boercker, Paul Lee, Edward H. Aifer, and Joseph G. Tischler*

**Supporting Information Content**:

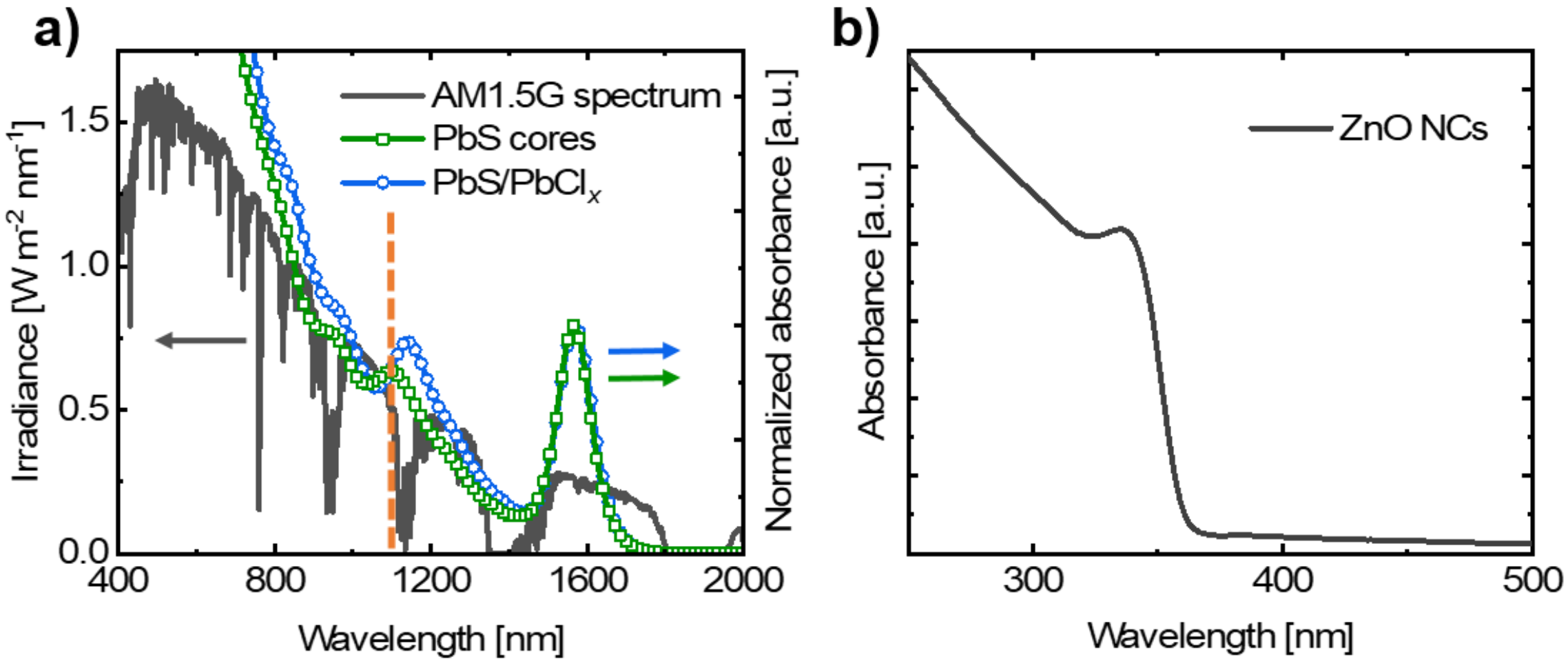

**Figure S1**. a) Solution absorbance spectra of the core-only (green squares) and core/shell (blue circles) PbS nanocrystal samples in tetrachloroethylene, overlaid with the AM1.5G solar spectrum. The dashed line represents the cutoff of the long pass filter at 1100 nm for the IR-illuminated *J*-*V* characterization. b) Solution absorbance spectrum of ZnO nanocrystals in chloroform.

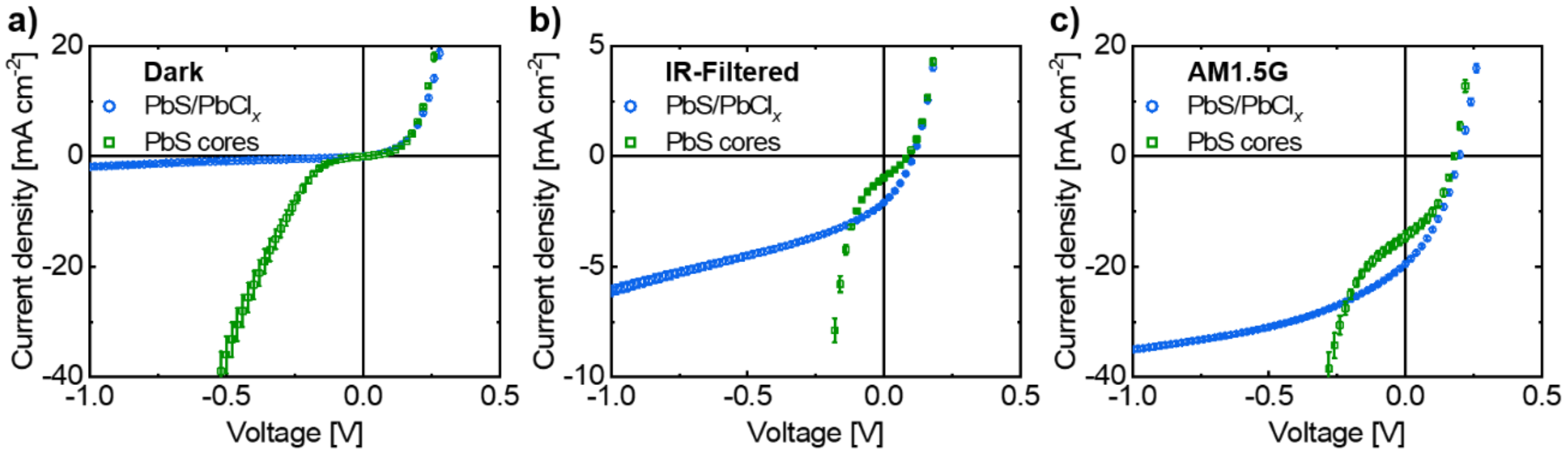

**Figure S2.** *J*-*V* characteristics of core-only (green squares) and core/shell (blue circles) devices a) in the dark, and under b) IR-filtered illumination (AM1.5G through a 1100 nm long pass filter) and c) full AM1.5G illumination. The active area of each device is 0.019 cm$^2$. Each curve is the average of 21 individual devices, and error bar uncertainties indicate their 95% confidence interval.

**Table S1.** Dark *J-V* characteristics for small-area core and core/shell PbS devices[a)]

| | | 0 V | | -0.25 V | | -0.5 V | |
|---|---|---|---|---|---|---|---|
| PbS Type | $R_S$ [b)] [Ω cm$^2$] | $J_{dark}$ [c)] [mA cm$^{-2}$] | $R_{SH}$ [d)] [Ω cm$^2$] | $J_{dark}$ [mA cm$^{-2}$] | $\Delta V/\Delta J$ [e)] [Ω cm$^2$] | $J_{dark}$ [mA cm$^{-2}$] | $\Delta V/\Delta J$ [Ω cm$^2$] |
| Core | 3.8 | -2.3x10$^{-4}$ | 230 | -8.4 | 11 | -36 | 7.0 |
| Core/Shell | 5.6 | -3.1x10$^{-4}$ | 300 | -0.48 | 720 | -0.84 | 640 |

[a)]Data corresponds to the dark *J-V* characteristics from Figure 1 of the main text, for detectors with active areas of 0.019 cm$^2$; [b)]($R_S$) is the series resistance calculated from $\Delta V/\Delta J$ at $V$ = 0.25 V; [c)]$J_{dark}$ is the dark current density; [d)]($R_{SH}$) is the shunt resistance calculated from $\Delta V/\Delta J$ at, and $V$ = 0 V; [e)] $\Delta V/\Delta J$ is the differential resistance at the given reverse bias voltage.

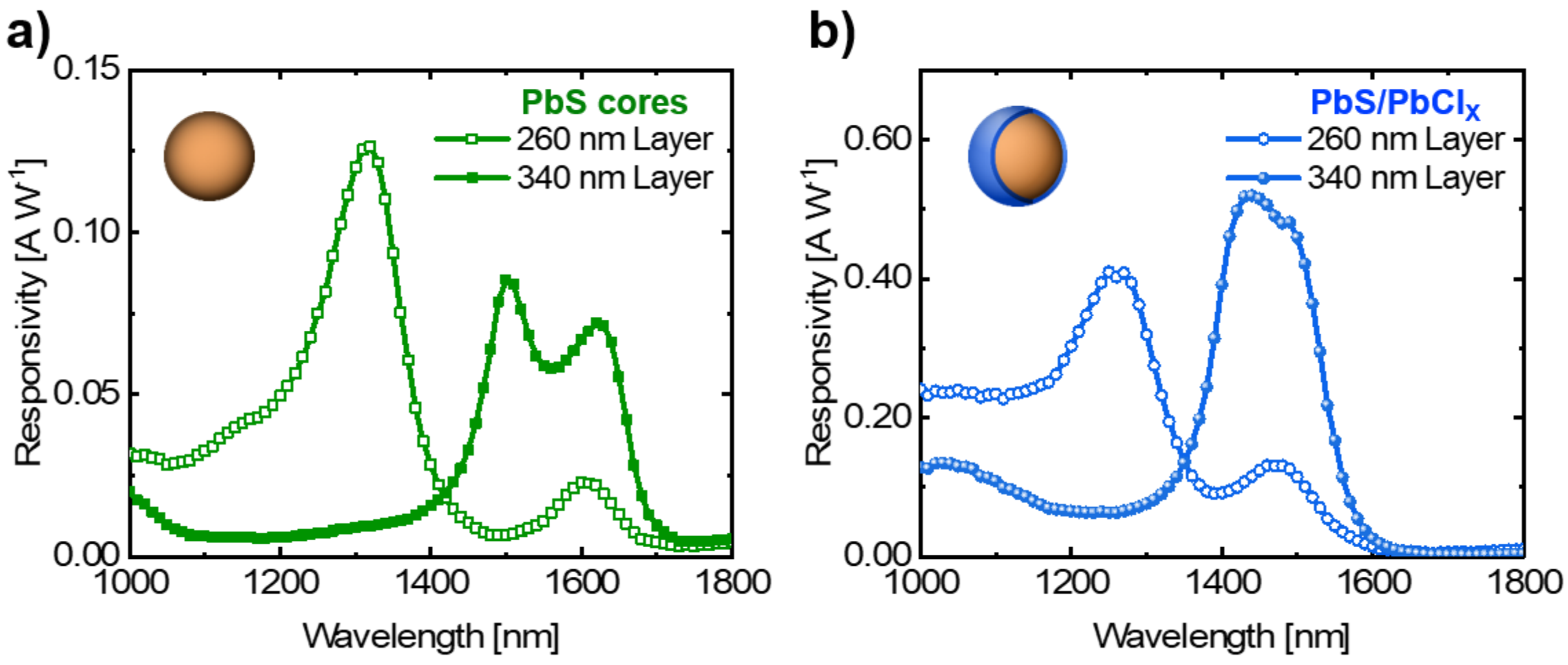

**Figure S3.** Zero-bias responsivity spectra for a) core-only and b) core/shell nanocrystal devices with active layer thickness of 260 nm (open symbols) and 340 nm (closed symbols). This corresponds to the EQE data shown in Figure 2 of the main text. All devices were fabricated with a 0.125 cm$^2$ active area. Responsivity ($R$) is determined using the equation: $R$ = EQE x ($qh^{-1}\nu^{-1}$), where $q$ is the charge of an electron, $h$ is Planck's constant, and ν is the frequency of the incident photon.

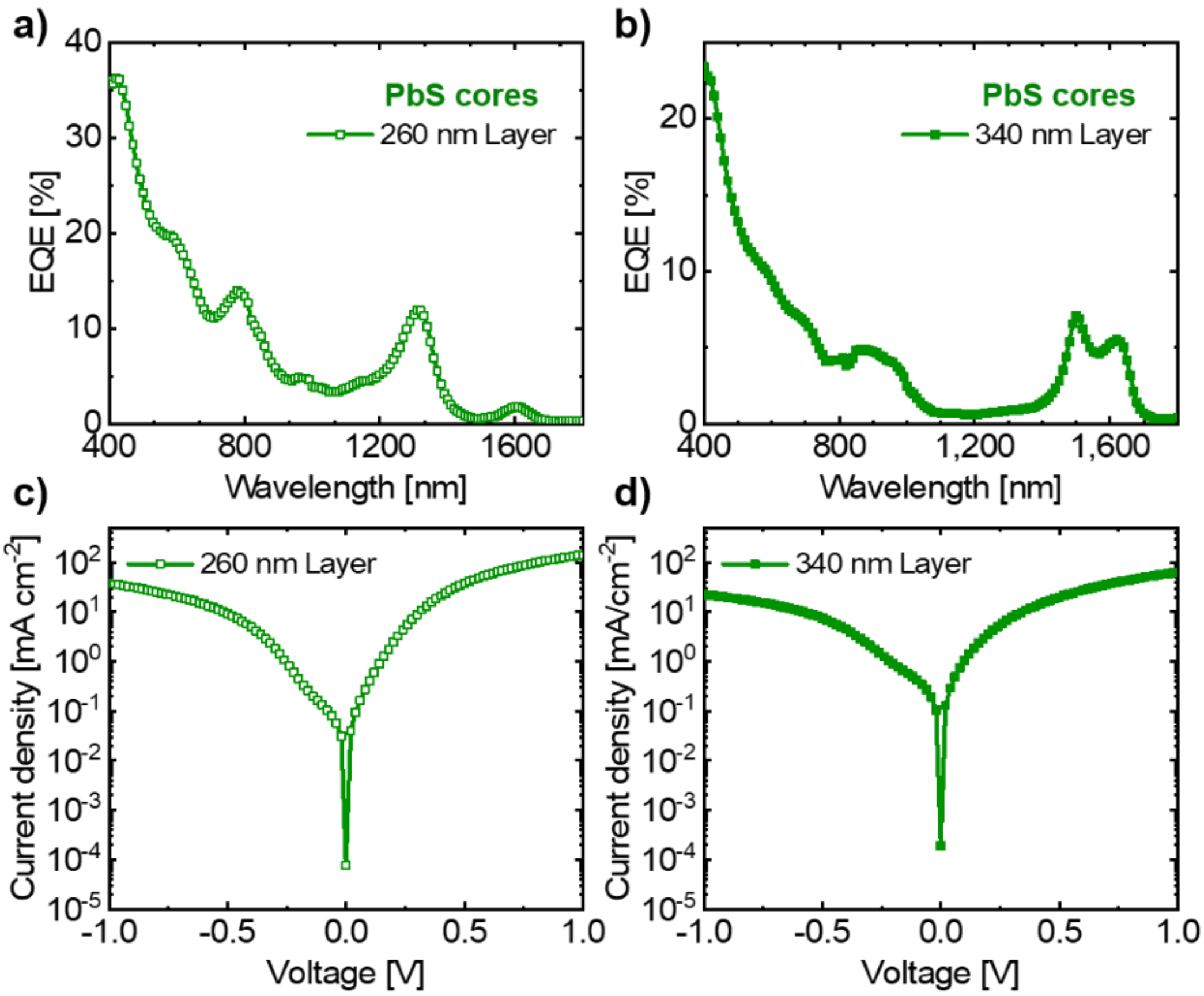

**Figure S4.** a,b) Zero-bias EQE spectra of PbS core-only devices with c,d) corresponding dark *J-V* characteristics for active layer thicknesses of a,c) 260 nm, and b,d) 340 nm, measured on large-area 0.125 $cm^2$ devices.

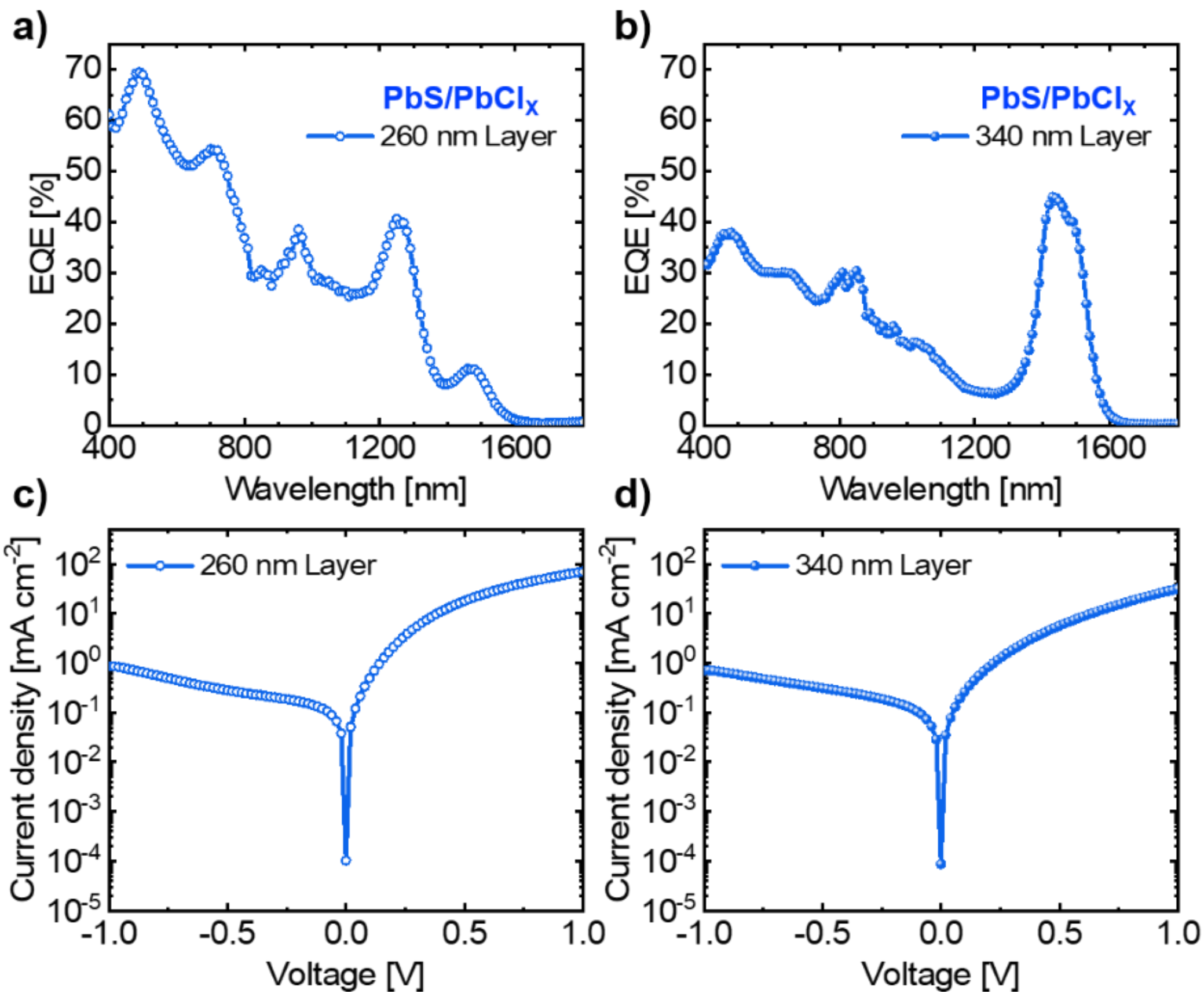

**Figure S5.** a,b) Full EQE spectra of core/shell devices with c,d) corresponding dark *J*-*V* characteristics for active layer thicknesses of a,c) 260 nm and b,d) 340 nm, measured on large-area 0.125 $cm^2$ devices.

**Table S2.** Comparison of device $J_{SC}$ values (measured vs calculated)

| PbS Type | Illumination[a)] | $J_{SC}$ Calculated[b)] [mA cm$^{-2}$] | $J_{SC}$ Measured[c)] [mA cm$^{-2}$] |
|---|---|---|---|
| Core/Shell | AM1.5G | -22.2 | -20.1 |
| Core/Shell | IR-filtered | -2.9 | -2.1 |
| Core | AM1.5G | -7.0 | -14.5 |
| Core | IR-filtered | -0.7 | -1.0 |

[a)]The IR source is 100 mW cm$^{-2}$ AM1.5G filtered past the band edge of c-Si ($\lambda$ > 1100 nm);

[b)]Calculated $J_{SC}$ values are determined from integration of the EQE spectra with the AM1.5G spectrum, for the 0.125 cm$^2$ active area devices from Figure 2 of the main text; [c)]Measured $J_{SC}$ values are those from Table 1 of the main text with 0.019 cm$^2$ active areas.

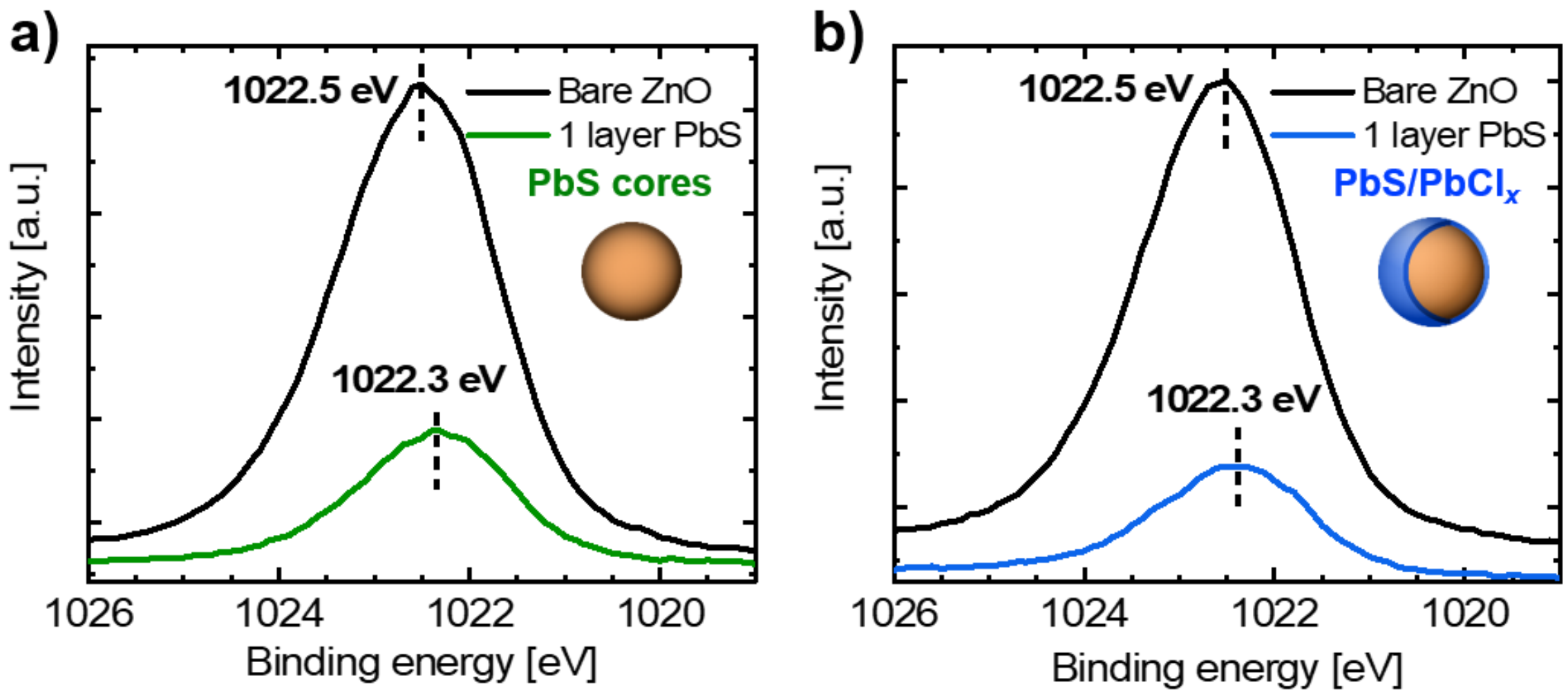

**Figure S6.** XPS core level spectra of Zn 2p$_{3/2}$ levels as deposited (0 PbS Layers) and with one layer of PbS nanocrystals deposited (1 PbS Layer, ~10 nm) for a) core-only and b) core/shell nanocrystals. There is an initial shift in both cases of 0.2 eV, suggestive of band bending at the interface. The Zn 2p$_{3/2}$ signal is lost after deposition of subsequent PbS layers due to exceeding the escape depth of electrons in XPS (5-10 nanometers).

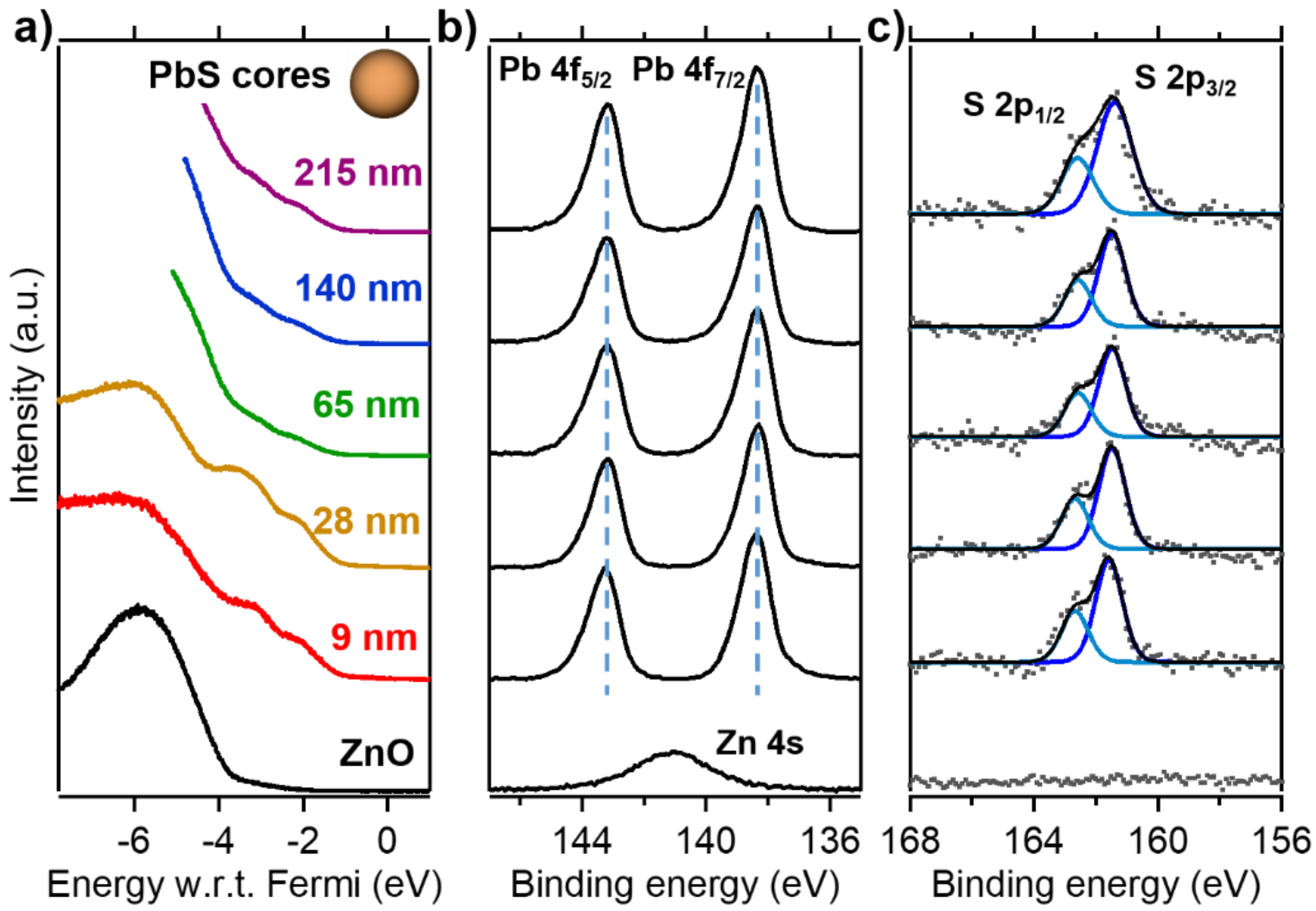

**Figure S7**. a) High kinetic energy edge UPS spectra for the successive deposition of core-only PbS plotted with respect to the system Fermi level, with corresponding XPS spectra for b) Zn 4s, Pb $4f_{5/2}$, Pb $4f_{7/2}$, and c) S $2p_{1/2}$ and S $2p_{3/2}$.

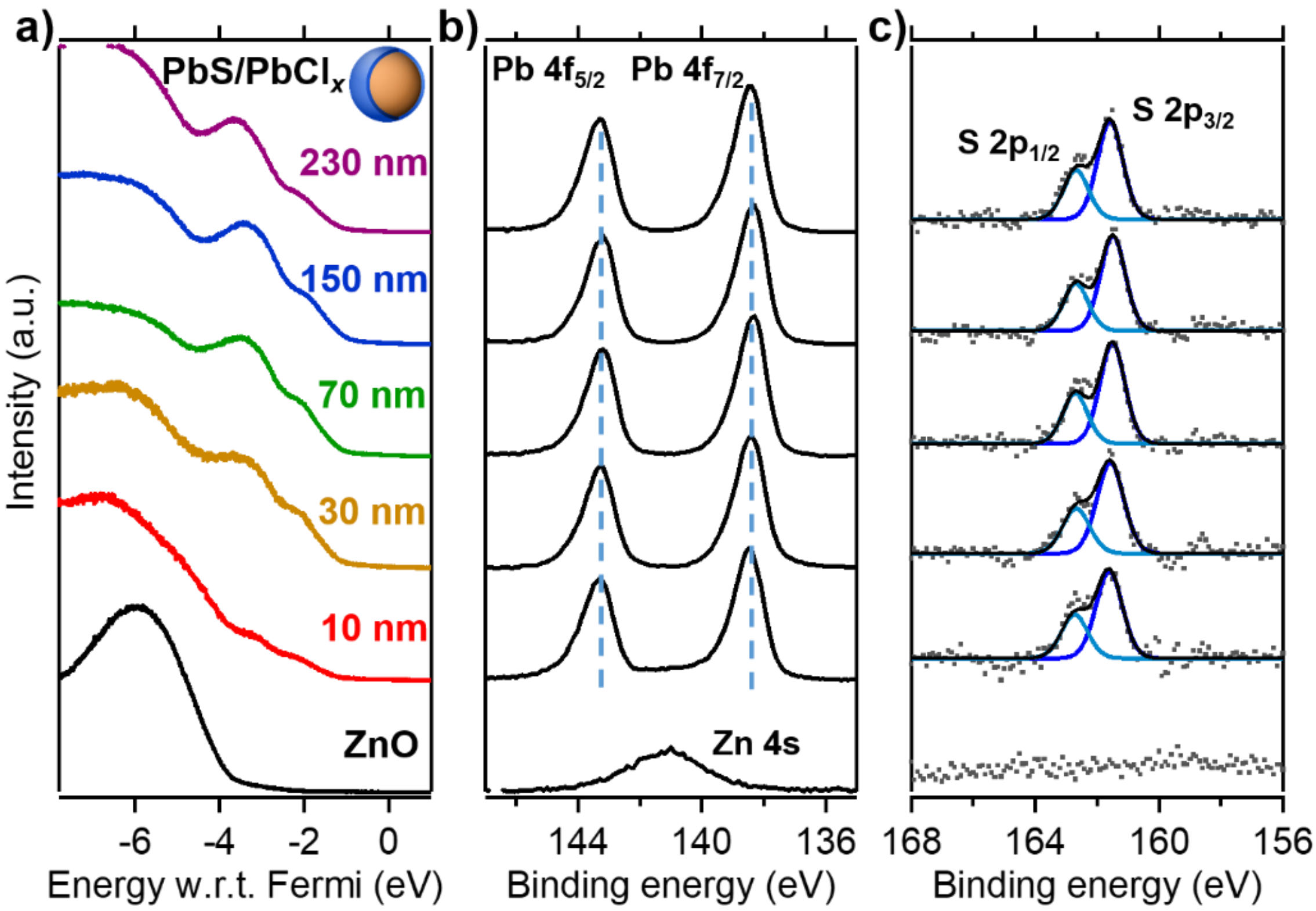

**Figure S8.** a) High kinetic energy edge UPS spectra for the successive deposition of core/shell nanocrystals plotted with respect to the system Fermi level, with corresponding XPS spectra for b) Zn 4s, Pb 4f$_{5/2}$, Pb 4f$_{7/2}$, and c) S 2p$_{1/2}$ and S 2p$_{3/2}$.

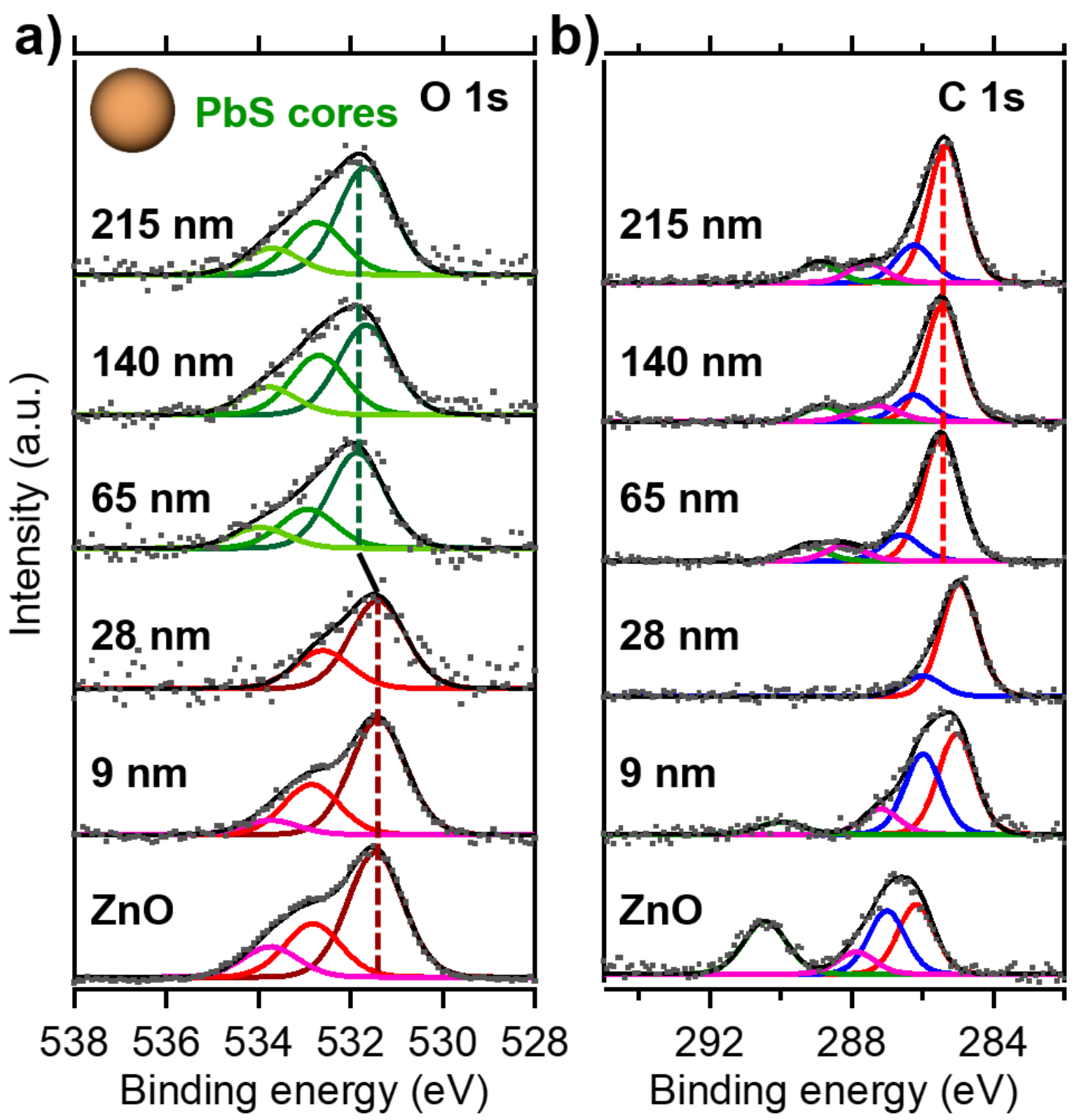

**Figure S9**. XPS core level spectra for a) O 1s and b) C 1s for the successive deposition of core-only nanocrystals. Here we observe a significant change in the oxygen and carbon spectra between the second and third measured layers (~30, and ~70 nm PbS, respectively) as the ZnO signal is suppressed, indicating incomplete removal of the native oleic acid ligands in the bulk of the core films. These changes correspond to the dipole-induced vacuum level shift of the work function observed between these layers (see Figure 3a,c, top panels).

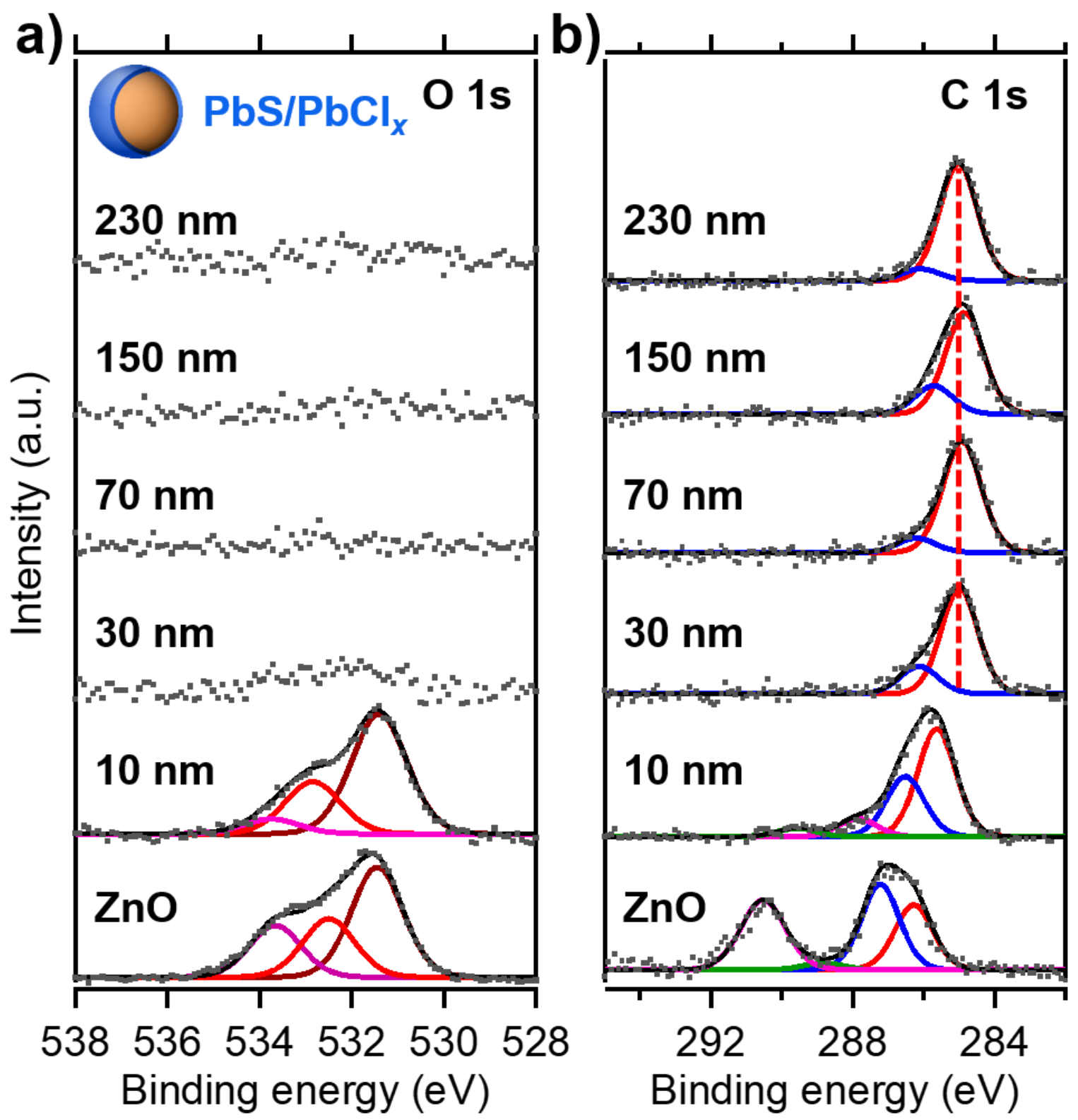

**Figure S10.** XPS core level spectra for a) O 1s and b) C 1s for the successive deposition of core/shell nanocrystals. Here we observe a distinct absence of oxygen following suppression of the ZnO signal, and the carbon peak remains consistent throughout the thickness of the film. This data indicates near complete removal of the native organic ligands present in the core/shell nanocrystals following iodide ligand exchange.